\documentclass[aps,prd,amsmath,superscriptaddress,%
               nobibnotes,nofootinbib,preprintnumbers,
               onecolumn,12pt,tightenlines]{revtex4}

\usepackage{graphicx}

\usepackage{ifthen}

\newcommand{\eg}{\textit{e.g.}}
\newcommand{\etal}{\textit{et~al.}}

\newcommand{\refeqn}[2][eqn:]{Eqn.~(\ref{#1#2})}

\newcommand{\reffig}[2][fig:]{Figure~\ref{#1#2}}
\newcommand{\Reffig}[2][fig:]{Figure~\ref{#1#2}}
\newcommand{\refsec}[2][sec:]{Section~\ref{#1#2}} 

\newcommand{\ifmulticol}[2]{%
  \ifthenelse{\lengthtest{1.9\columnwidth<\textwidth}}{#1}{#2}%
}

\newcommand{\scfigwidth}{%
  \ifdim1.9\columnwidth<\textwidth%
    1.00\columnwidth\else0.60\columnwidth\fi%
}

\newcommand{\widefigwidth}{%
  \ifdim1.9\columnwidth<\textwidth%
    1.00\columnwidth\else1.00\columnwidth\fi%
}

\newcommand{\insertfig}[1]{%
    \includegraphics[keepaspectratio,width=\scfigwidth,
                     height=0.35\textheight]{#1}
}

\newcommand{\insertwidefig}[1]{%
    \includegraphics[keepaspectratio,width=\widefigwidth,
                     height=0.35\textheight]{#1}
}

\newcommand{\insertdoublefig}[2]{%
    \includegraphics[keepaspectratio,width=0.49\textwidth,
                     height=0.35\textheight]{#1}
    \hspace{\stretch{1}}
    \includegraphics[keepaspectratio,width=0.49\textwidth,
                     height=0.35\textheight]{#2}
}

\newcommand{\gae}{%
  \ensuremath{\lower 2pt \hbox{%
    $\, \buildrel {\scriptstyle >}\over {\scriptstyle \sim}\,$}%
    }%
  }
\newcommand{\lae}{%
  \ensuremath{\lower 2pt \hbox{%
    $\, \buildrel {\scriptstyle <}\over {\scriptstyle \sim}\,$}%
    }%
  }

\newcommand{\Leff}{\ensuremath{\mathcal{L}_{\textrm{eff}}}}
\newcommand{\Enr}{\ensuremath{E_{nr}}}

\newcommand{\Eest}{\ensuremath{E^{\prime}}}
\newcommand{\Snr}{\ensuremath{S_{nr}}}
\newcommand{\See}{\ensuremath{S_{ee}}}
\newcommand{\Sone}{\ensuremath{S1}}
\newcommand{\Stwo}{\ensuremath{S2}}
\newcommand{\etaSone}{\ensuremath{\eta_{S1}}}
\newcommand{\Soneave}{\ensuremath{\langle\Sone\rangle}}

\begin{document}




\title{XENON10/100 dark matter constraints in comparison with CoGeNT
       and DAMA: examining the \Leff\ dependence}

\author{Christopher Savage}
\email[]{savage@fysik.su.se}
\affiliation{
 The Oskar Klein Centre for Cosmoparticle Physics,
 Department of Physics,
 Stockholm University,
 AlbaNova,
 SE-10691 Stockholm, Sweden}

\author{Graciela Gelmini}
\email[]{gelmini@physics.ucla.edu}
\affiliation{
 Department of Physics and Astronomy,
 UCLA,
 430 Portola Plaza,
 Los Angeles, CA 90095, USA}

\author{Paolo Gondolo}
\email[]{paolo@physics.utah.edu}
\affiliation{
 Department of Physics,
 University of Utah,
 115 S 1400 E \#201
 Salt Lake City, UT 84112, USA}
\altaffiliation[Also at]{
 Korea Institute for Advanced Study,
 Seoul 130-722, Korea
 }

\author{Katherine Freese}
\email[]{ktfreese@umich.edu}
\affiliation{
 Michigan Center for Theoretical Physics,
 Department of Physics,
 University of Michigan,
 Ann Arbor, MI 48109, USA}

\date{\today}



\begin{abstract} 

\vspace{0.2in}
\begin{center}
  \textbf{Abstract}
\end{center}

We consider the compatibility of DAMA/LIBRA, CoGeNT, XENON10 and
XENON100 results for spin-independent (SI) dark matter Weakly
Interacting Massive Particles (WIMPs), particularly at low masses
($\sim$10~GeV), assuming a standard dark matter halo.  The
XENON bounds depend on the scintillation efficiency factor \Leff\ for
which there is considerable uncertainty. Thus we consider various
extrapolations for \Leff\ at low energy.  With the \Leff\ measurements
we consider, XENON100 results are found to be insensitive to
the low energy extrapolation. We find the strongest bounds are from
XENON10, rather than XENON100, due to the lower energy threshold.
For reasonable choices of \Leff\ and for the case of SI elastic
scattering, XENON10 is incompatible with the DAMA/LIBRA 3$\sigma$
region and severely constrains the 7-12~GeV WIMP mass region of
interest published by the CoGeNT collaboration.

\end{abstract} 

\maketitle


\section{\label{sec:Intro} Introduction}

The nature of the dark matter that comprises a quarter of the Universe
is one of the big unanswered questions in astrophysics and particle
physics.  Perhaps the best motivated candidates are Weakly Interacting
Massive Particles (WIMPs) which have weakly interacting cross sections
and masses in the GeV--10~TeV range.  In recent months, there have been
new data releases from many experiments that have engendered a great
deal of excitement. Of particular interest is a low mass region
$\sim 10$~GeV which at first sight seems to be compatible with a number
of different experiments.  The goal of this paper is to examine some of
the issues regarding the question of such compatibility.   In this
paper, we will restrict ourselves here to considering WIMPs with
spin-independent (SI) interactions.  Spin-dependent and mixed couplings
will be examined in a future work.

It was initially the DAMA/NAI experiment
\cite{Bernabei:2010mq}, looking for annual modulation
\cite{Drukier:1986tm,Freese:1987wu} of a WIMP signal, that found such a
possible low mass region. WIMPs with SI interactions in the mass
range 5--9~GeV were found to be compatible with the DAMA/NaI
results and all negative results from other searches that existed at
the time~\cite{Gelmini:2004gm,Gondolo:2005hh}. The situation changed
after the publication of the first DAMA/LIBRA results
\cite{Bernabei:2008yi} (see \eg\ 
Ref.~\cite{Savage:2008er} and reference therein). 
For SI interactions, Ref.~\cite{Savage:2008er} found that the best fit
DAMA regions were ruled out to the 3$\sigma$  C.L.
But Ref.~\cite{Savage:2008er} also found that for WIMP masses of
$\sim$8~GeV, some
parameters outside these regions still yielded a moderately reasonable
fit to the DAMA data and were compatible with all 90\% C.L.\ upper
limits from negative searches, when ion channeling in the DAMA
experiment as understood at the time was included  (see
\refsec{Channeling} below).  The strongest bounds at the time
came from CDMS \cite{Ahmed:2008eu} and XENON10 \cite{Angle:2007uj}.
Since then many new data sets have been released and a reexamination
of the light WIMP region is now necessary.  We will
focus on the following three factors: a possible dark matter signal for
7-12~GeV WIMP's found by the CoGeNT collaboration~\cite{Aalseth:2010vx};
the existence of new better upper limits, in particular the bounds set
by the XENON10~\cite{Angle:2009xb} and XENON100~\cite{Aprile:2010um}
collaborations; and the recognition that the effect of channeling in
NaI(Tl) crystal is less important than previously
assumed~\cite{Bozorgnia:2010xy} (see \refsec{Channeling}).

CDMS \cite{CDMSII} has released its full data set, with tighter bounds
and two unexplained events at low energy that may be compatible with
background.  CDMS constraints will be included but are not a focus of
this paper.  Our focus is on elastic scattering of spin-independent
WIMPs from a standard Maxwellian halo; recent examinations comparing
experimental studies in this case include
\cite{Bottino:2009km,Kopp:2009qt,Fitzpatrick:2010em,Chang:2010yk,
Petriello:2008jj,Bottino:2008mf,Chang:2008xa,Hooper:2008cf}.

Just prior to our paper being submitted, a revised version of
Ref.~\cite{Andreas:2010dz} appeared that added an examination of
XENON100 and \Leff\ in the context of scalar WIMPs.
The halo model parameters and \Leff\ models used in that paper differ
somewhat from ours, but the results are qualitatively similar.

The interpretation of the XENON10 and XENON100 results requires the
ability to reliably reconstruct the nuclear recoil energy from the
observed signal. This depends on the  scintillation efficiency factor
\Leff\ for which there is considerable uncertainty at low energies
(see \refsec{Leff}). A large part of this paper is devoted to
examining the \Leff\ dependence of the two XENON constraints. 

In this paper we focus on comparing the following experimental results:
the combined modulation signal \cite{Bernabei:2010mq} as well as the
total rate \cite{Bernabei:2008yi} from DAMA/NaI and DAMA/LIBRA;
the combined CDMS 5-tower results \cite{Ahmed:2008eu,CDMSII};
the recent first results from XENON100 \cite{Aprile:2010um};
and the older but lower threshold XENON10 reanalysis results
\cite{Angle:2009xb}.
Constraints for CDMS, XENON10, and XENON100 are determined using
the Maximum Gap method\footnote{
    For zero observed events (as in the case of XENON100), the Maximum
    Gap (MG) method provides an identical constraint as that produced
    by a Poisson limit based on the total number of events.  When
    events are observed, the MG method provides better constraints than
    the Poisson case as the former takes the energy spectrum into
    account, whereas the latter does not.  The use of MG is of
    particular importance for XENON10 as the 13 observed events have
    energies that are inconsistent with the spectrum expected for a
    light elastically scattering WIMP.
    }
\cite{Yellin:2002xd}, while the parameters
compatible with DAMA are determined via the goodness-of-fit of their
observed modulation signal with the theoretically expected signal.
Details of these statistical analyses may be found in
Ref.~\cite{Savage:2008er}.

For the two XENON experiments, we assume the energy resolution is
primarily limited by a Poisson distribution in the small number
of photoelectrons (PE) expected at low recoil energies.
Interactions in the liquid Xenon comprising the XENON10 and XENON100
detectors give rise to a prompt scintillation signal (\Sone) followed
by a delayed secondary scintillation signal (\Stwo). The quantities
\Sone\ and \Stwo\ are discussed in the following section
and the various thresholds and data cuts are described in \eg\ 
Ref.~\cite{Angle:2009xb}.
The efficiencies for XENON10 and XENON100 are taken from
Refs.~\cite{Angle:2009xb} and~\cite{Aprile:2010um}, respectively.
However, for XENON10, the \Sone\ peak finding efficiency
factor \etaSone, which was not included in Ref.~\cite{Angle:2009xb}
(where it was not particularly relevant), must also be taken into
account.
For this \etaSone\ factor, we take the more conservative of the two
cases found in Ref.~\cite{Sorensen:2010HEFTI}.  

A second issue must also be taken into account.
As explained in detail below, as the recoil energy decreases, so do the
average \Sone\ and \Stwo\ signals.  At low enough energies, a sizable
fraction of the events may fail to produce enough \Stwo\ signal and/or
to fall in the proper $\log(\Stwo/\Sone)$ range (the nuclear recoil band
cut), even if a fluctuation
gives a high enough \Sone.  At higher recoil energies, the relative
size of the fluctuations get smaller and the average \Stwo\ is too
high for any significant fraction of the \Stwo\ fluctuations to fall
below the \Stwo\ threshold.  However, somewhere below an \Sone\ average
of 1~photoelectron (PE) in XENON10, an event would need not only an upward
fluctuation in \Sone, but an upward fluctuation in \Stwo\ to pass both
thresholds and it would require particular ranges of fluctuations to
fall within the required range for $\log(\Stwo/\Sone)$. Thus,
to avoid issues with the \Stwo\ threshold and nuclear recoil band cuts,
we ignore recoil energies that give on average less than 1~PE in the
\Sone\ signal.
We find that, with Poisson fluctuations included, low energy recoils
in XENON10 tend to pass the various thresholds and cuts at a higher
rate than predicted by the efficiencies we use; these efficiencies can
thus be considered conservative over the range at which they are
applied.
Additional events passing these various cuts can arise from low energy
recoils that give an average \Sone\ signal \Soneave\ that
falls below our imposed cutoff of $\Soneave \ge 1.0$~PE.  Accounting
for these events will strengthen the XENON10 constraints at low WIMP
masses.  Our \Soneave\ cutoff is simply due to the efficiency being
unknown for these low energy recoils at this time.
We are examining these low energy recoil efficiencies for a future work,
though a comprehensive treatment of these efficiencies in XENON10 and
XENON100 has appeared in the meantime \cite{Sorensen:2010hq}.

In addition to constraints for the above experiments,
we also show the 7-12~GeV WIMP mass region suggested
by CoGeNT as an explanation for excess events seen at low energies
in their detector \cite{Aalseth:2010vx}.  We perform our own
statistical analyses of all data sets except CoGeNT,
for which we simply use their published region.

To allow for direct comparison, all the other experimental constraints
are determined using the same 600~km/s galactic halo escape velocity as
used by CoGeNT in their analysis \cite{Collar:2010pc}.
This escape velocity falls within the 90\% confidence interval of
498~km/s to 608~km/s that was determined by a recent analysis of high
velocity stars \cite{Smith:2006ym}, but is somewhat above the median
likelihood of 544~km/s that was found in that analysis.
Similarly, we use throughout the value 220~km/sec (used by CoGeNT) for
the rotation velocity of the Galactic disk in the vicinity of the Sun.
We note that new measurements suggest that the rotation velocity might
be higher, $254 \pm 16$~km/s \cite{Reid:2009nj}, the effect of which is
to shift the best fit in all experiments to slightly lower WIMP
masses \cite{Savage:2009mk}.

\textbf{Comparison with Version One of this paper:}
We have made two substantial changes, as discussed in the previous
paragraphs. First, we have included the effects of \etaSone, the \Sone\ 
peak finding efficiency factor.  With this factor included, the XENON10
constraints shift upwards in scattering cross-section by at most
$\sim$~$\times$1.7 at any given WIMP mass, not by several orders of
magnitude (as was speculated in Ref.~\cite{Collar:2010nx}).
Second, in the current version we do not include recoil energies which
yield on average less than 1.0~PE in the \Sone\ signal.
It is this second issue that is by far the bigger effect.
A more complete discussion can be found in the Appendix.
In addition, in the Appendix we respond to various critiques of our work.

\section{\label{sec:Issues} Experimental Issues: \Leff\ and Channeling}

In this section we discuss two important experimental issues.  First,
the scintillation factor in XENON \Leff\ is extremely important in
interpreting results and yet is not well known.  Second, the channeling
effect in DAMA/LIBRA, again not well known, may change the location of
the regions in WIMP parameter space that are compatible with the data.

\subsection{\label{sec:Leff} The \Leff\ Scintillation Efficiency Factor
            in XENON}

The interpretation of the XENON10 and XENON100 results requires the
ability to reliably reconstruct the nuclear recoil energy  from the
observed signal. This reconstruction depends on the  scintillation
efficiency factor \Leff\ for which there is considerable uncertainty
at low energies.  Here we discuss this factor and present three
models for \Leff\ at low energy.

Interactions in the liquid Xenon comprising the XENON10 and XENON100
detectors give rise to a prompt scintillation signal, \Sone, followed
by a delayed secondary scintillation signal, \Stwo.  The \Sone\ signal
arises from a rapid relaxation of excited Xenon states produced as a
result of the interaction.  The \Stwo\ signal arises from ionized
electrons also produced in the interaction; these drift through the
liquid xenon under an applied electric field, but once they reach the
liquid surface they are extracted into a xenon gas phase where they
emit proportional scintillation light.  The drift time of the
electrons causes this secondary scintillation (the \Stwo\ signal) to
be observed later than the \Sone\ signal, allowing both scintillation
signals to be measured separately.
The \Sone\ signal can be used to determine the energy of the
interaction, while the combination of both signals allows discrimination
between nuclear recoil events (possibly WIMP interactions) and
electron recoil events (necessarily background interactions). The ratio
of \Stwo\ to \Sone\ is much higher in the case of electron recoils than
in the case of nuclear recoils.

Interpretation of the XENON results requires the ability to reliably
reconstruct the nuclear recoil energy \Enr\ from the observed \Sone\ 
signal.
Calibration of the nuclear recoil energy dependence of \Sone\ 
often involves gauging the detector's response to electron recoils at
higher energies; parts of the detector's response (\eg\ the fraction of
scintillation photons that yield photoelectrons (PE) in the
photodetectors) are more easily determined in this case than with
nuclear recoils at lower energies.
Taking \Sone\ to be normalized to the number of PE,
\Sone\ and \Enr\ are related by an equation involving the higher
energy electron recoil calibrations:
\begin{equation} \label{eqn:S1}
  \Sone = (\Snr / \See) \, \Leff(\Enr) \, L_y \, \Enr \, .
\end{equation}
Here, $L_y$ is the light yield in PE/keVee for 122~keVee $\gamma$-rays%
\footnote{
    The unit ``keVee'' refers to the electron-equivalent energy in keV,
    the amount of energy in an electron recoil event that would
    produce a given scintillation signal in the detector (whether or
    not the scintillation was, in fact, produced by an electron recoil).
    The unit ``keVnr'' refers to the nuclear recoil energy in keV.
    }.
\Leff(\Enr) is the scintillation efficiency of
nuclear recoils relative to 122~keVee $\gamma$-rays in zero electric
field;  this factor is a function of the nuclear recoil energy.  Since
there is an applied electric field in the experiment, which reduces the
scintillation yield by quickly removing charged particles from the
original interaction region,
two additional factors must be taken into account:
\See\ and \Snr\ are the suppression in the scintillation yield for
electronic and nuclear recoils, respectively, due to the presence of
the electric field in the detector volume.
The quantities \See, \Snr, and $L_y$ are detector dependent; \Leff\ is
not.

\begin{figure}
  \insertfig{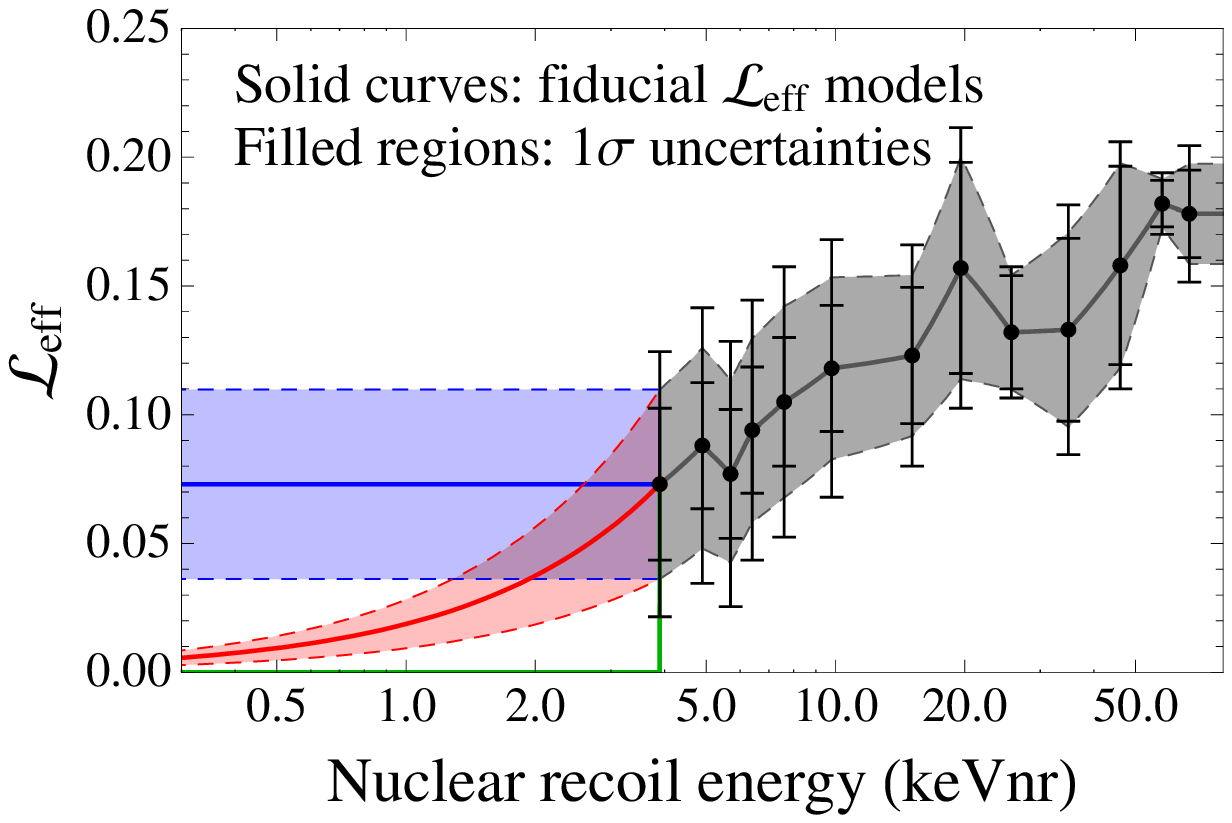}
  \caption[\Leff\ vs.\ recoil energy]{
    \Leff\ as a function of recoil energy.
    The points correspond to the measurements of Manzur
    \etal\ \cite{Manzur:2009hp} with statistical and systematic errors
    in \Leff\ as indicated (uncertainties in recoil energy not shown).
    Solid curves show the fiducial \Leff\ dependence used in this work.
    Filled regions/dashed curves indicate the 1$\sigma$ variation in
    the \Leff\ dependence.
    At 3.9~keVnr (the lowest energy data point),
    $\Leff = 0.073_{-0.025-0.026}^{+0.034+0.018}
    \approx 0.073 \pm 0.037$.
    At lower energies three cases are examined:
    constant \Leff\ (blue) at the above value,
    \Leff\ falling linearly to zero at zero energy (red),
    and \Leff\ equal to zero (green).
    Above 3.9~keVnr, the gray curve and region are used in all cases.
    Note linear relationships appear curved in the figure due to the
    logarithmic scaling.
    }
  \label{fig:Leff}
\end{figure}

Recent comments have drawn attention to the role that \Leff\
determinations play in setting experimental constraints for Xenon-based
detectors \cite{Collar:2010gg,Collaboration:2010er,Collar:2010gd}.
A variety of \Leff\ measurements have been made over the years
\cite{Arneodo:2000vc,Bernabei:2001pz,Akimov:2001pb,Aprile:2005mt,
Chepel:2006yv,Sorensen:2008ec,Aprile:2008rc,Manzur:2009hp}, but
limited statistics and systematics issues have so far prevented a
clear picture from emerging as to the behavior of \Leff\ at low recoil
energies.
There are two primary issues in debate:
(1) Which of the \Leff\ measurements should be used as a basis for
analyzing direct detection results? and (2) Measurements of \Leff\ 
have only been made at energies above some minimum; what is the
behavior of \Leff\ at low energies, where no measurements have as yet
been made?
For the first issue, the XENON100 collaboration has chosen to use a
global fit to multiple \Leff\ measurements in their analysis, whereas
Ref.~\cite{Collar:2010gg} suggests that the recent measurements by
Manzur \etal\ \cite{Manzur:2009hp} should be used;
in both cases, \Leff\ measurements are based upon fixed energy
neutron scatters.
We do not contribute to the debate as to which
\Leff\ data sets are most appropriate; however, in the interest of
examining the most conservative XENON constraints, we use the
Manzur \etal\ data alone in our analyses.  

The choice of \Leff\ measurements to use in the XENON analyses has a
significant impact on the resulting constraints for low WIMP masses.
The Manzur \etal\ data yield the lowest values for \Leff\ 
among the fixed-energy neutron scatter measurements, implying the
highest recoil energy thresholds and therefore the lowest sensitivity
for the XENON detectors to low mass WIMPs (which generate only low
energy recoils).  The Manzur \etal\ data is shown in \reffig{Leff}.

A comment is in order about the lower ZEPLIN-III \Leff\ measurement
\cite{Lebedenko:2008gb} represented as a band in Fig.~1 of
\cite{Collar:2010gg}.
ZEPLIN fits a nonlinear \Leff\ model
to their broad spectrum nuclear recoil calibration data to obtain
\Leff\ curves that were used in their analysis.  These fits suggest a
constant \Leff\ at recoil energies above $\sim$30~keVnr, with \Leff\ 
sharply falling at energies below $\sim$20~keVnr and approaching zero
at $\sim$7-8~keVnr; see Figure~15 of Ref.~\cite{Lebedenko:2008gb} and
the accompanying text.
Thus the suggestion has been made by nonmembers of the ZEPLIN team
\cite{Collar:2010nx} that \Leff\ should be taken to be zero below
$\sim$8~keVnr as a conservative model of \Leff.  This \Leff\ model
would yield significantly weaker XENON constraints relative to what we
have referred to as conservative models based on the Manzur \etal\ 
measurements of \Leff\ \cite{Manzur:2009hp}.
However, the members of the ZEPLIN experiment themselves do not
advocate their fits as being an indicator of \Leff\ behavior at recoil
energies below $\sim$ 8~keVnr \cite{Sumner:2010pc}.  In addition, the
dependence of these curves on statistical and systematic uncertainties
has not been fully determined, where these uncertainties can
significantly impact the lowest recoil energy portion of their fits.
The ZEPLIN-III dark matter analysis is, in fact, mainly insensitive to
the low recoil energy portion of their \Leff\ curves.
Further discussion of the ZEPLIN data can be found in the Appendix.
Moreover, as explained in detail in the
Appendix, below the recoil energies of 7~keVnr, 
the Manzur \etal\ measurements are incompatible with $\Leff = 0$ at far
more than the 3$\sigma$ level. We consider the Manzur \etal\ 
measurements more reliable than the ZEPLIN-III estimate of \Leff\ at
low energies.

The second issue in debate is how \Leff\ behaves at energies below where
measurements have been made.  Most \Leff\ measurements are at recoil
energies above 5~keVnr; Manzur \etal\ have a measurement at
$3.9 \pm 0.9$~keVnr.  The \Leff\ behavior below these energies is
unclear from an experimental and theoretical standpoint, at least at
the precision necessary for use in a WIMP constraint analysis.  The
XENON collaboration has suggested that \Leff\ measurements are
consistent with \Leff\ being effectively constant at low recoil
energies, at least at energies where recoils may contribute to their
signal \cite{Aprile:2010um,Aprile:2008rc,Sorensen:2008ec}.
Various \Leff\ measurements are also consistent with an \Leff\ that
decreases as one goes to lower recoil energies; see \eg\ Sect.~V of
Ref.~\cite{Manzur:2009hp} which provides a theoretically motivated
empirical model of such a decreasing \Leff.
Furthermore, Ref.~\cite{Collar:2010gg} states that
``the mechanisms behind the generation of \textit{any} significant
amount of scintillation are still unknown and may simply be absent at
the few keVnr level.'' Given this uncertainty we use three different
extrapolations of \Leff\ at low energies: constant, decreasing as one
goes to lower recoil energies, or just zero.

\begin{figure*}
  \insertdoublefig{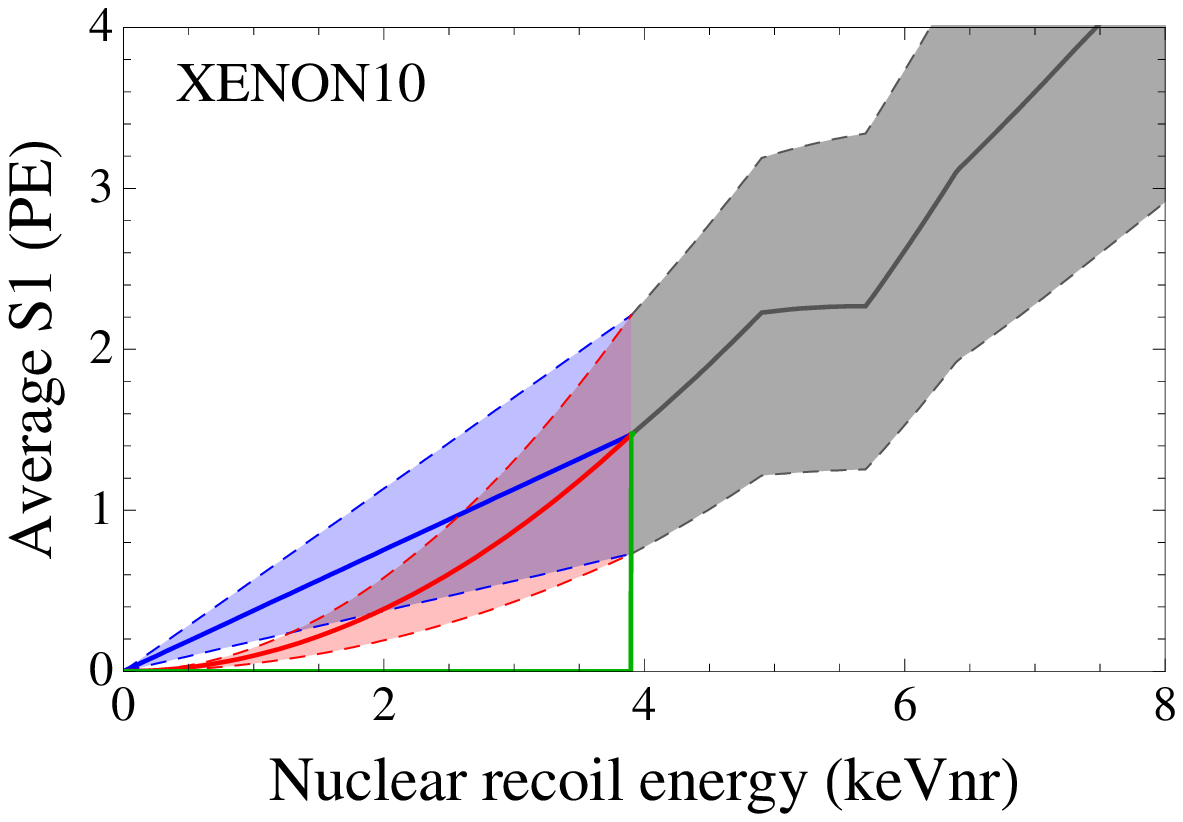}{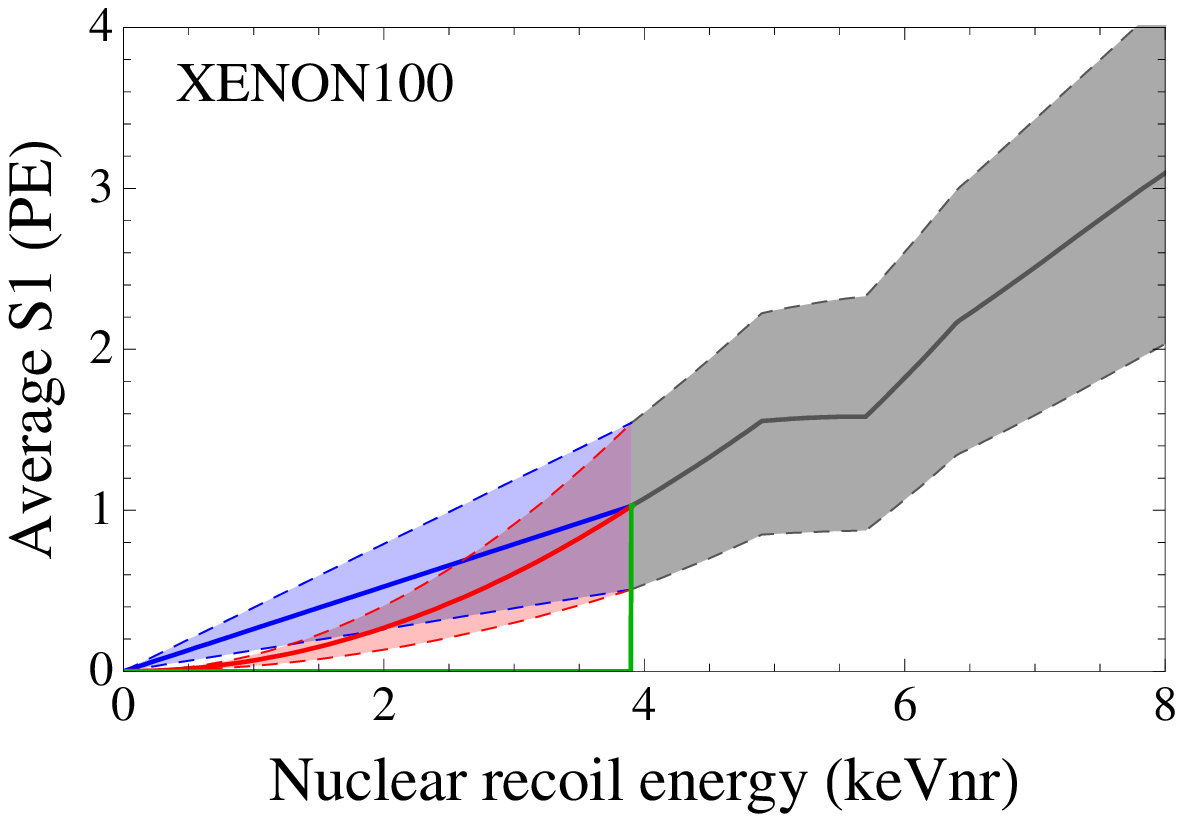}
  \caption[\Leff\ vs.\ recoil energy]{
    The average \Sone\ signal as a function of nuclear recoil energy
    for XENON10 (left) and XENON100 (right).
    Curves and regions correspond to the \Leff\ models shown in
    \reffig{Leff}.
    }
  \label{fig:S1Enr}
\end{figure*}

We choose as our fiducial \Leff\ model a piecewise linear interpolation
between the central Manzur \etal\ values at their measured energies,
shown in \reffig{Leff}.
In addition, we will also examine similarly constructed \Leff\ models
using the 1$\sigma$ uncertainties in the Manzur \etal\ measurements%
\footnote{
    The statistical and systematic errors in the measured \Leff\ are
    added in quadrature with the upper and lower uncertainties
    averaged; uncertainties in the corresponding recoil energies for
    those measurements have been neglected.}.
The choice of linear interpolation vs.\ a quadratic interpolation or
spline fit to the \Leff\ points has a negligle impact on the generated
constraints compared to that from the 1$\sigma$ variations in the
\Leff\ points themselves.
Below recoil energies of 3.9~keVnr, the lowest Manzur \etal\ 
measurement\footnote{
    Manzur \etal\ measure $\Leff = 0.073_{-0.025-0.026}^{+0.034+0.018}
    \approx 0.073 \pm 0.037$
    at 3.9~keVnr, where the two errors in the first case are the
    statistical and systematic uncertainties, respectively, and the
    second case is the combined uncertainty as described in the
    previous footnote.
},
we examine three behaviors for \Leff, also shown in \reffig{Leff}:
(1) a constant \Leff,
(2) an \Leff\ that goes linearly to zero at zero recoil energy, and
(3) an \Leff\ that is strictly zero.
Even if the scintillation goes to zero at some low but finite recoil
energy, there is no reason to expect this to occur above
$\sim$2-3~keVnr; the measurements of \Leff\ provide no indication of
an abrupt (rather than gradual) falling of \Leff\ at energies just
below where the measurements exist.  As such, the third case is
perhaps unrealistically conservative, but never-the-less provides the
most conservative case.  In addition, the use of this case will allow
us to examine the contribution of low energy recoils in generating
constraints.
The average \Sone\ signals as a function of the nuclear recoil energy
\Enr\ that correspond to these \Leff\ models are shown in \reffig{S1Enr}
for XENON10 and XENON100.

In the interest of examining the most conservative XENON constraints,
we base all three cases on the data from Manzur
\etal\ \cite{Manzur:2009hp}.  Of the existing data sets, the Manzur
\etal\ data yield the lowest values for \Leff, implying higher recoil
energy thresholds for the XENON experiments, and thereby reducing the
sensitivity of XENON to low mass WIMPs.

\subsection{\label{sec:Channeling} Channeling Effects in DAMA/LIBRA}

The channeling effect is of crucial importance when considering the
compatibility of DAMA with other experimental results as this effect
has the potential to significantly alter the WIMP masses and
cross-sections which are compatible with the DAMA modulation signal.

Channeling and blocking effects in crystals refer to the orientation
dependence of charged ion penetration in crystals. In the ``channeling
effect,'' ions incident upon a crystal along symmetry axes and planes
suffer a series of small-angle scattering that maintain them in the
open``channels'' in between the rows or planes of lattice atoms and
thus penetrate much further into the crystal than in other directions.
Channeled incident ions do not get close to lattice sites, where they
would be deflected at large angles, and they lose energy almost
exclusively into electrons. The ``blocking effect'' consists in a
reduction of the flux of ions originating in lattice sites along
symmetry axes and planes, creating what is called a ``blocking dip''
in the flux of ions exiting from a thin enough crystal as a function of
the exit angle with respect to a particular symmetry axis or plane.
The potential importance of the channeling effect for direct dark
matter detection was first pointed out by
H.~Sekiya \etal~\cite{japanese} and subsequently for NaI(Tl) by
Drobyshevski~\cite{Drobyshevski:2007zj} and by the DAMA
collaboration~\cite{Bernabei:2007hw}.  When Na or I ions recoiling
after a collision with a dark matter WIMP are channeled, their
quenching factor\footnote{
    The quenching factor $Q$ is the ratio of ionization or
    scintillation produced by a nuclear recoil event in a crystal
    relative to that produced in an electron recoil event of the
    same energy.  This is analogous to the \Leff\ factor in liquid
    Xenon and is likewise used to reconstruct the nuclear recoil
    energy from the observed ionization/scintillation of an event.
    }
is approximately $Q=1$ instead of $Q_I=0.09$ and
$Q_{Na}=0.3$, since they give their energy to electrons. The DAMA
collaboration~\cite{Bernabei:2007hw} estimated the fraction of
channeled recoils and found it to be large for low recoiling energies
in the keV range. Using this evaluation of the channeling fraction, the
regions in cross-section versus mass of acceptable WIMP models in
agreement with the DAMA data were found to be considerably shifted
towards lower WIMP masses and cross-sections.

\begin{figure}
  \insertfig{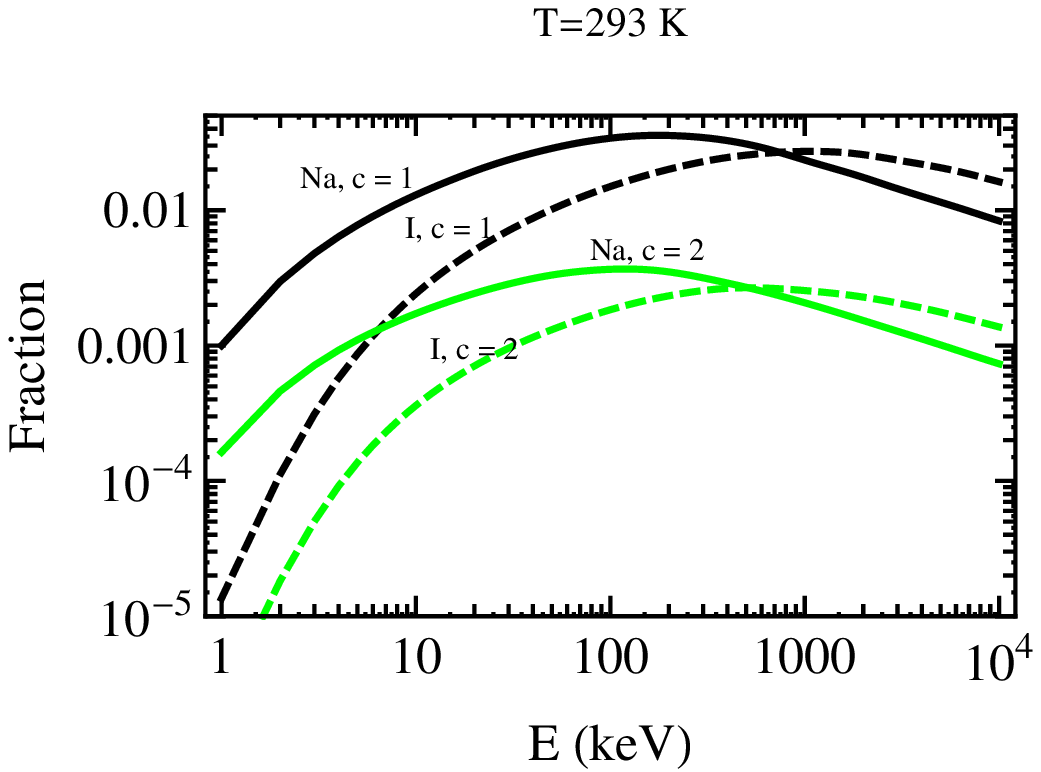}
  \caption[Channeling fractions]{
    Upper bounds to the channeling fraction at a temperature of 293~K
    for Na (solid lines) and I (dashed lines) recoiling ions in a NaI
    crystal for two different models of the temperature effect in the
    lattice parameterized with $c=1$ (black) and $c=2$ (green or gray).
    No dechanneling processes are taken here into account. To be 
    conservative, in this paper we will use the $c=1$ results presented
    here, as they yield the largest change in the DAMA compatible
    regions of parameter space relative to the no-channeling case.
    This figure is reproduced from Ref.~\cite{Bozorgnia:2010xy},
    where further details may be found.
    }
  \label{fig:Frac-RoomT}
\end{figure}

However, the DAMA calculation of the channeling fraction did not take
into account that the recoiling lattice ions start initially from
lattice sites (or very close to them) and, therefore, blocking effects
are important. In fact, as argued originally by
Lindhard~\cite{Lindhard:1965}, in a perfect lattice and in the absence
of energy-loss processes, the probability of a particle starting from a
lattice site to be channeled would be zero. The argument uses
statistical mechanics in which the probability of particle paths
related by time-reversal is the same. In a perfectly rigid lattice, the
fraction of channeled recoils would, in fact, be zero. However, the
atoms in a crystal are actually vibrating about their equilibrium
positions in the lattice.  It is this displacement from equilibrium
that allows for a non-zero channeling
probability of recoiling ions. The vibration amplitude increases with
the temperature, thus the effect is temperature dependent: in general the
channeling fraction increases with temperature. 

Upper bounds to the recoiling channeling fractions in NaI(Tl) crystals
at 20$^{\circ}$C were obtained in Ref.~\cite{Bozorgnia:2010xy}, using
analytic models of channeling developed since the 1960's, when
channeling was discovered (see for example
Refs.~\cite{Lindhard:1965,Gemmell:1974ub,Hobler} and references
therein). These upper bounds on the channeling fractions were obtained
with temperature effects taken into account not only through the
vibrations of the colliding nucleus but also in the lattice. The latter
depend on the parameter $c$ (see Ref.~\cite{Bozorgnia:2010xy} for
details) which in the relevant literature is found to be a number
between 1 and 2, with 1 giving the largest channeling fractions
(see \reffig{Frac-RoomT}, reproduced from Ref.~\cite{Bozorgnia:2010xy}).

The fractions shown in \reffig{Frac-RoomT} are also an upper bound in
that no dechanneling mechanism has been taken into account to compute
them. The collisions with Tl impurities would take channeled ions out
of their channel, and this process is not included (see
Ref.~\cite{Bozorgnia:2010xy} for further explanations).

\section{\label{sec:Results} Results and Discussion}

\begin{figure*}
  \insertwidefig{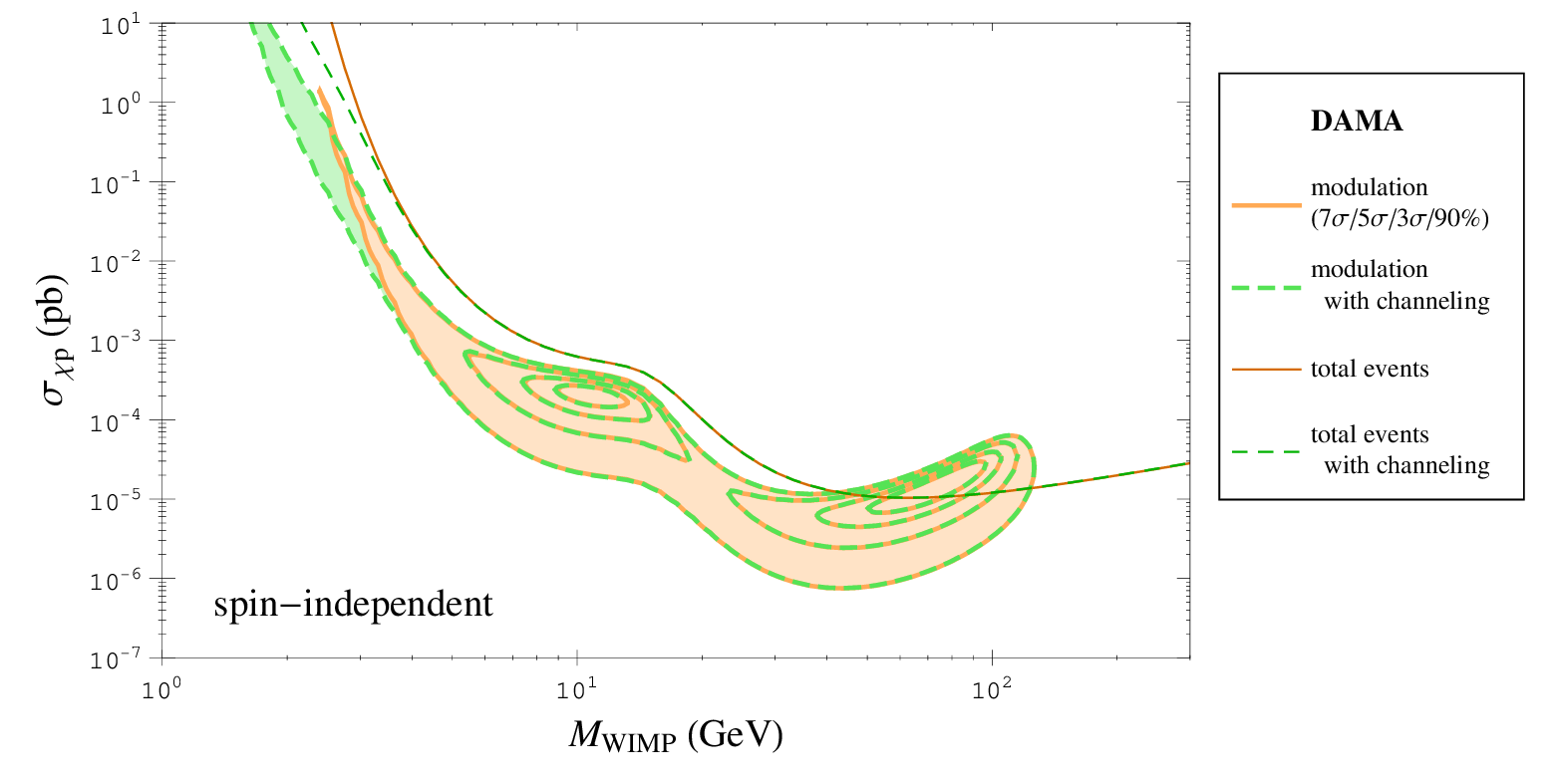}
  \caption[DAMA Channeling]{
    WIMP masses and spin-independent (SI) cross-sections compatible
    with the DAMA modulation signal and total number of events,
    determined with (dashed green) and without (solid orange) the
    channeling effect included.  The largest channeling fractions shown
    in \reffig{Frac-RoomT} (taken from Ref.~\cite{Bozorgnia:2010xy})
    are used here for the channeling case.
    Comparing the cases with or without channeling, we find
    negligible difference in the DAMA modulation
    regions at the 90\%, 3$\sigma$, and 5$\sigma$ levels; only
    the 7$\sigma$ contours differ and only for WIMP masses below 
    4~GeV.
    The lower and higher mass DAMA regions correspond to parameters
    where the modulation signals arise from scattering predominantly
    off of Na and I, respectively.
    }
  \label{fig:DAMA}
\end{figure*}

In \reffig{DAMA}, we show the WIMP masses and
SI cross-sections compatible with the DAMA modulation signal both
with and without channeling included; contours are shown for regions
compatible at the 7$\sigma$, 5$\sigma$, 3$\sigma$, and 90\% level
(in order from larger to smaller regions).
For the channeling, we use the largest channeling fractions
shown in \reffig{Frac-RoomT} as they provide the largest
potential effect on the DAMA constraints.
\Reffig{DAMA} shows that even in this case there is negligible
difference between the channeling and
non-channeling scenarios except for regions incompatible with DAMA
at greater than the 5$\sigma$ level.  Even in these cases, the
difference lies only at WIMP masses below 4~GeV and at relatively high
SI cross-sections.  As channeling is a negligible effect, we do not
further include it.

Compared to our previous analysis in ~\cite{Savage:2008er},
the current study takes advantage of additional recently released
DAMA data.  The effect of the additional data has been to sharpen
the regions in parameter space that match the data. 
For example, at $5\sigma$, there are now two completely separate
regions (peaked at different WIMP masses) that were previously joined.
In our current work we also display a $7\sigma$
contour in which the two regions are again connected.
We remind the reader that we are using the goodness-of-fit
statistic described in detail in Ref.~\cite{Savage:2008er}.

\begin{figure*}
  \insertwidefig{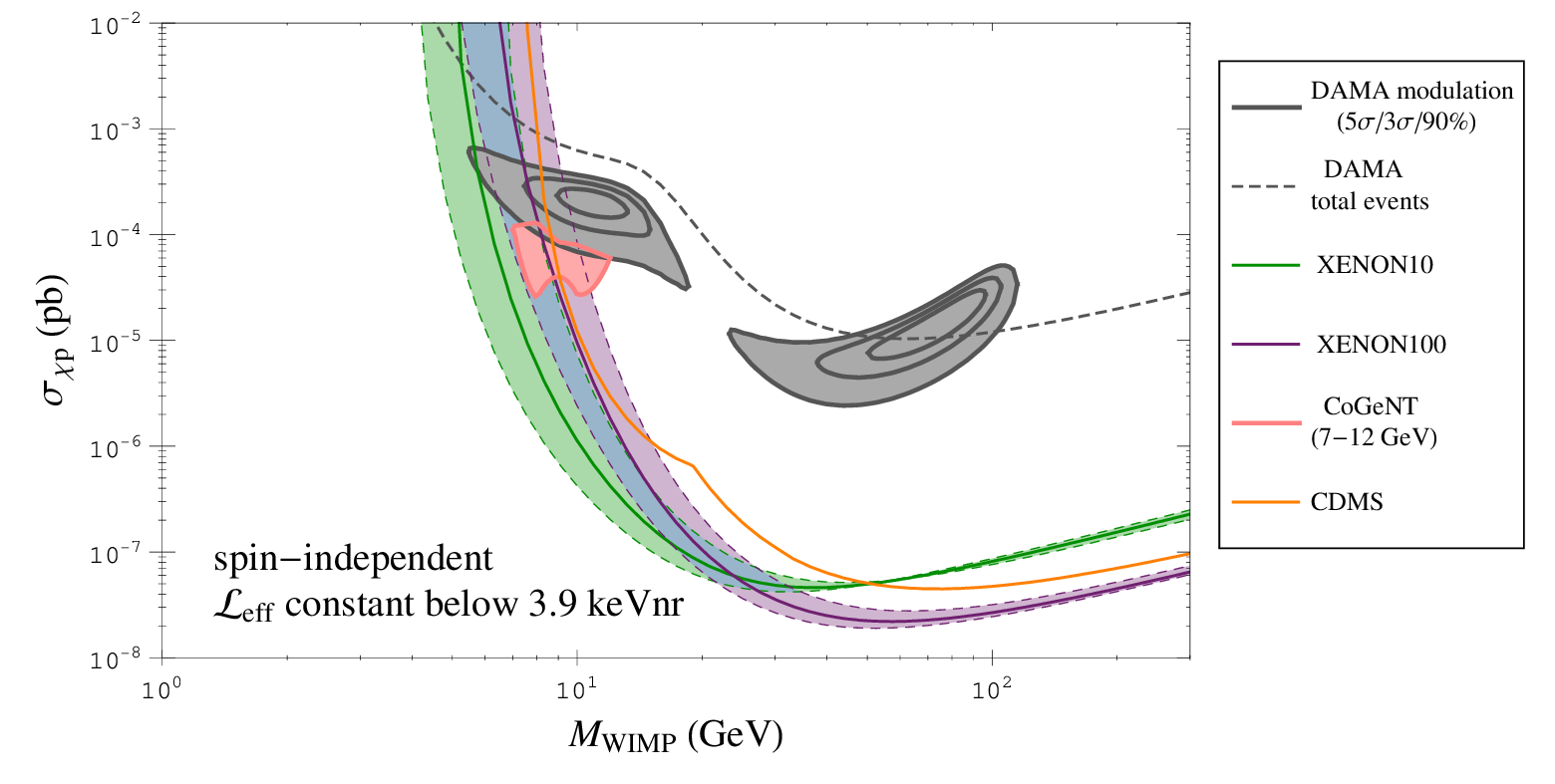}
  \caption[Flat \Leff]{
    XENON10 (green) and XENON100 (purple) 90\% C.L.\ constraints for a
    constant \Leff\ at recoil energies below 3.9~keVnr.
    The solid curves are the constraints using the central values
    of \Leff\ as described in the text; dashed curves and lighter
    filled regions indicate how these 90\% constraints vary with the
    1$\sigma$ uncertainties in \Leff.
    The blue region indicates an overlap between the XENON10 (green)
    and XENON100 (purple) 1$\sigma$ regions.
    Also shown are the CDMS constraint (orange curve), DAMA modulation
    compatible regions (gray contours/region), and the CoGeNT
    7-12~GeV region (pink contour/region).
    The lower and higher mass DAMA regions correspond to parameters
    where the modulation signals arise from scattering predominantly
    off of Na and I, respectively.
    }
  \label{fig:ConstantLeff}
\end{figure*}

\begin{figure*}
  \insertwidefig{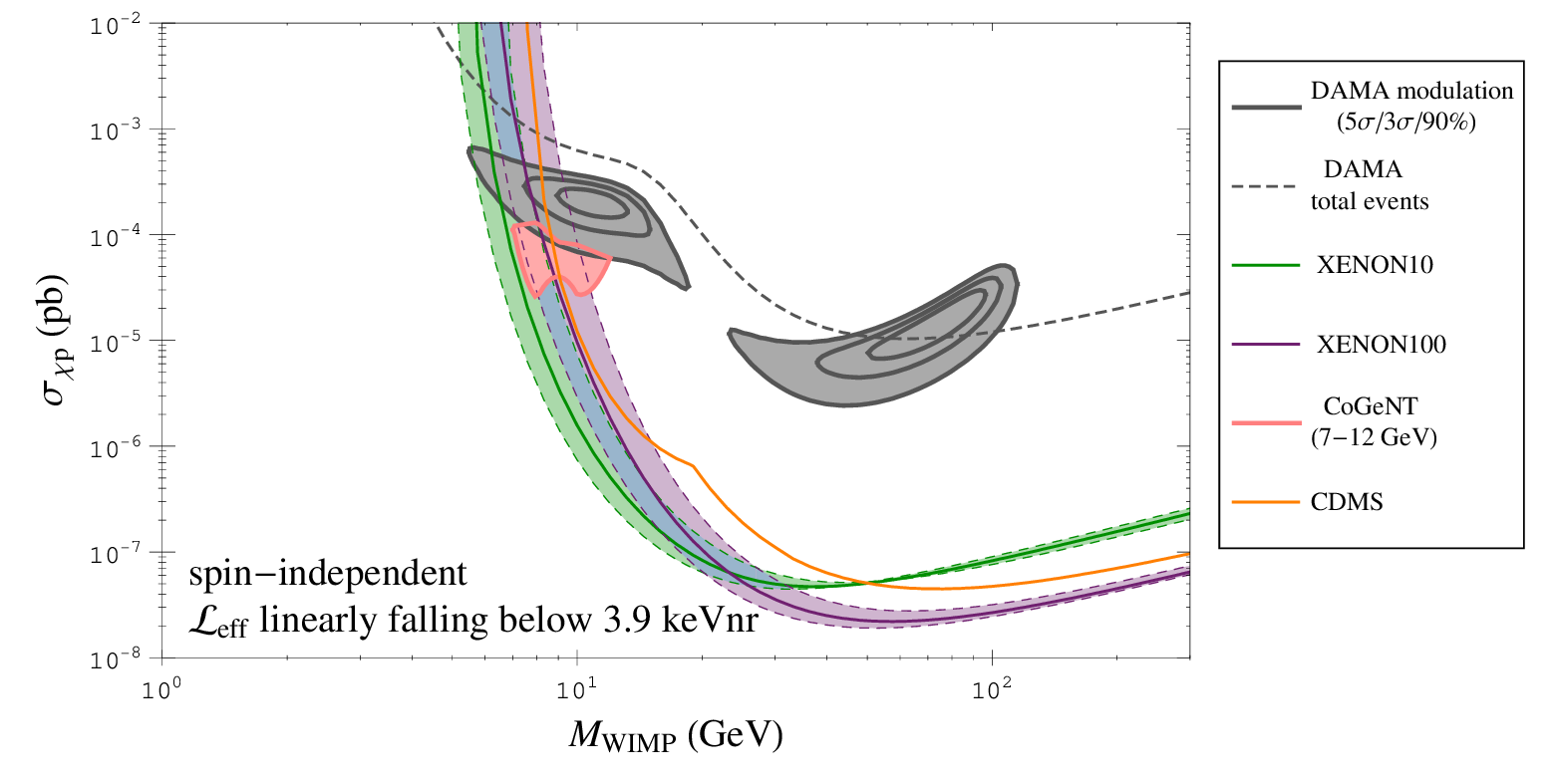}
  \caption[Flat \Leff]{
    Same as \reffig{ConstantLeff}, but taking \Leff\ to fall linearly
    to zero for recoil energies below 3.9~keVnr.
    }
  \label{fig:FallingLeff}
\end{figure*}

\begin{figure*}
  \insertwidefig{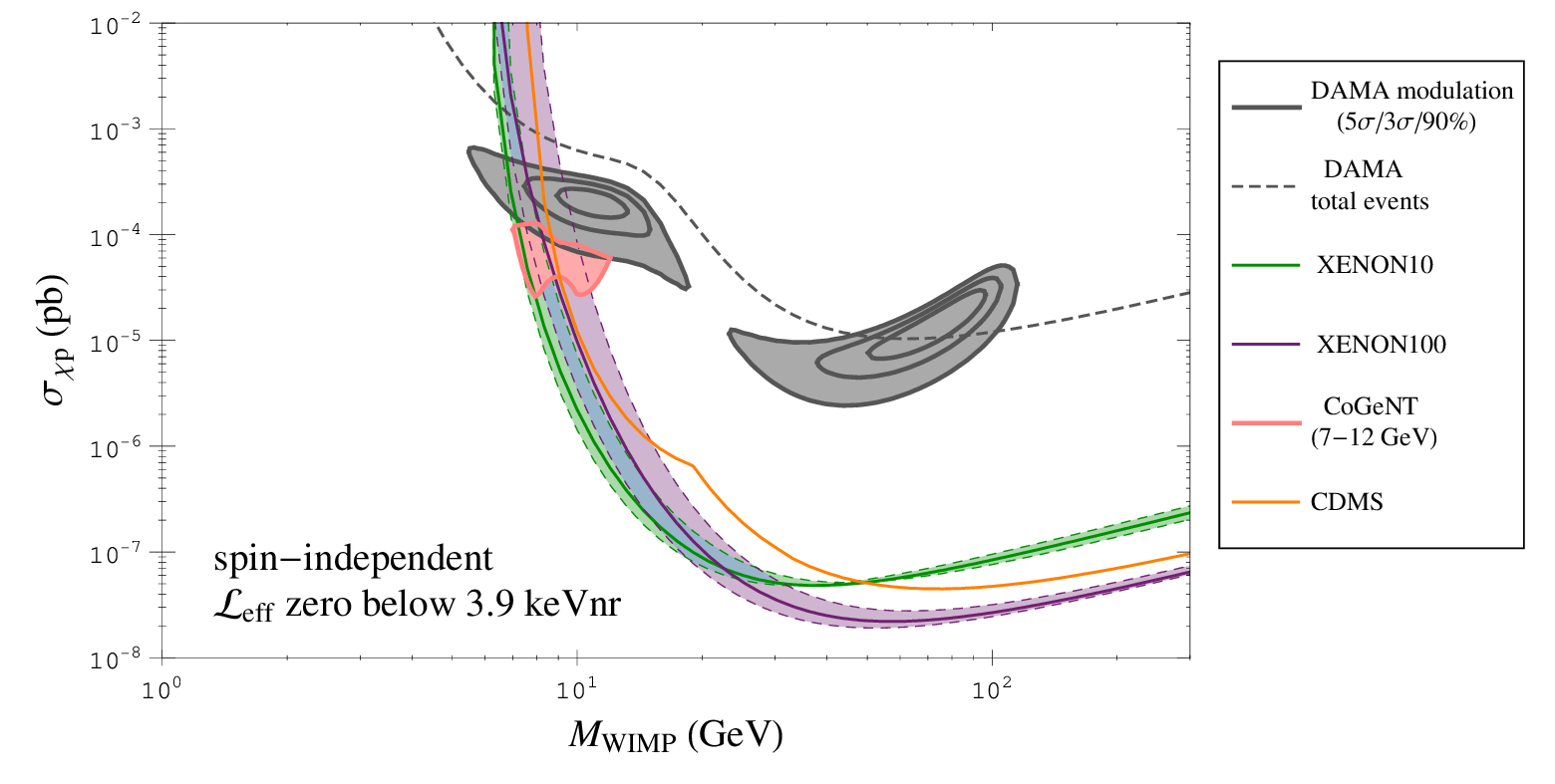}
  \caption[Flat \Leff]{
    Same as \reffig{ConstantLeff}, but taking \Leff\ to be zero for
    recoil energies below 3.9~keVnr.
    }
  \label{fig:ZeroLeff}
\end{figure*}

Our main results are shown in
Figs.~(\ref{fig:ConstantLeff})-(\ref{fig:ZeroLeff}),
corresponding to the three cases for the behavior of \Leff\ at low
recoil energies.  The solid gray contours indicate the WIMP parameters
compatible with the DAMA modulation within the 5$\sigma$, 3$\sigma$,
and 90\% level; the 5$\sigma$ DAMA region is also shaded light gray.
The (filled) pink contour corresponds to the 7-12~GeV WIMP mass
region suggested by CoGeNT (we reiterate that we have not reanalyzed
their data and simply display their published region here).
CDMS, DAMA (total events), XENON10,
and XENON100 curves indicate regions for which the WIMP parameters
are excluded at the 90\% level (the parameters above these curves
are excluded).  The solid green region for XENON10
and solid purple region for XENON100 do \textit{not} indicate regions
compatible within a given level (as opposed to the DAMA and CoGeNT
regions); they instead indicate how the 90\% exclusion
constraints vary with the 1$\sigma$ level uncertainties in the
\Leff\ measurements.
Overlapping XENON10 and XENON100 1$\sigma$ regions are shown in blue.

For the fiducial (central value) \Leff\ model in the case where it is 
constant below 3.9~keVnr, shown in \reffig{ConstantLeff}, the XENON100
constraint excludes all of the DAMA 3$\sigma$
region, but only the portion of the CoGeNT region with WIMP masses
above 9~GeV.  We note that, because we use only the Manzur \etal\ 
\Leff\ data \cite{Manzur:2009hp}, this constraint is weaker than that
presented by XENON100 \cite{Aprile:2010um}.  If the 1$\sigma$
uncertainties in \Leff\ are included,
XENON100 could exclude nearly all of the DAMA 5$\sigma$ region and
the entire CoGeNT region, if the largest
value of \Leff\ in the 1$\sigma$ region is taken.
On the other hand, it might exclude only the CoGeNT region above
11~GeV and not even all of the DAMA 90\% region, if the lowest
value of \Leff\ in the 1$\sigma$ region is used.  However, the CDMS
constraint, unaffected by the issues with \Leff, constrains the
same CoGeNT region as the fiducial XENON100 case here, with a slighter
weaker constraint on the DAMA region (incompatible with the DAMA
2$\sigma$ region, not shown).

For the fiducial (central value) \Leff\ model in the case where it
falls linearly to
zero at zero recoil energy, shown in \reffig{FallingLeff}, the XENON100
constraint again excludes nearly all of the DAMA 3$\sigma$ region and
the portion of the CoGeNT region with WIMP masses above 9~GeV.  The
1$\sigma$ variations in the \Leff\ measurements also yield a similar
variation in the XENON100 constraint as they did in the constant
\Leff\ case.
The most extreme case, taking \Leff\ to be zero below 3.9~keVnr, yields
similar XENON100 constraints as the other two cases, as seen in
\reffig{ZeroLeff},
although the constraint using the 1$\sigma$ upper values of \Leff\ 
does not quite exclude the full CoGeNT region, leaving a narrow window
at WIMP masses of 7-8~GeV.  It should be emphasized, however, that the
linearly falling \Leff\ case is already conservative and taking
\Leff\ to be zero below 3.9~keVnr is perhaps unrealistically
conservative.

The XENON100 constraints are nearly identical in the DAMA and CoGeNT
regions for all three cases of low energy \Leff\ behavior.  In fact,
the constraints based on the central and 1$\sigma$ lower values of
\Leff\ \textit{are} identical; only when using the upper 1$\sigma$
\Leff\ values do the constraints differ.
There are two main reasons for the similarity among the constraints:
(1) the imposed $\Soneave \ge 1.0$~PE cutoff and
(2) the small potential contribution from recoil events with energies
below 3.9~keVnr where the \Leff\ models differ.
As can be seen in \reffig{S1Enr}, a recoil energy of 3.9~keVnr
yields an average \Sone\ signal of 1.0~PE in XENON100 when using any
of the three fiducial \Leff\ models.  With the \Soneave\ cutoff, there
is no contribution from recoils at energies below 3.9~keVnr where
the fiducial \Leff\ models differ; thus, these constraints are
identical.  When using the 1$\sigma$ lower values of \Leff, no
recoils below 5.9~keVnr are included, so the lesser constraining
portion of the 1$\sigma$ XENON100 constraint bands shown in the
figures are likewise identical.
On the other hand, when using the 1$\sigma$ upper values of \Leff,
the $\Soneave \ge 1.0$~PE cutoff corresponds to recoil energies of
2.5, 3.1, and 3.9~keVnr for the constant, linearly falling, and
zero low energy \Leff\ models, respectively.
In this case, low energy recoils contribute to the constraints.
However, these low energy recoils can make only a small contribution
to the observed signal, as will be discussed below.

\begin{figure*}
  \insertwidefig{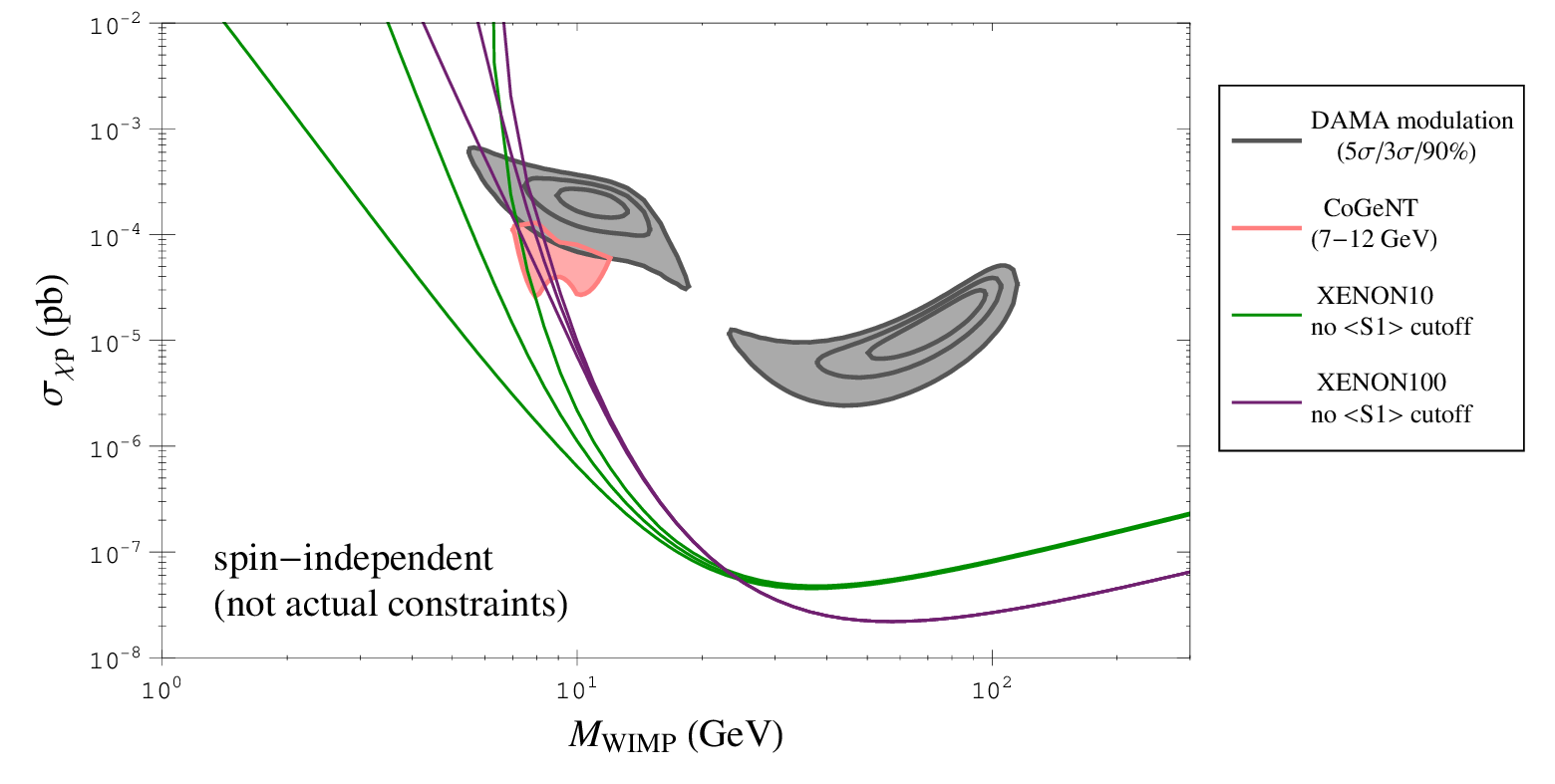}
  \caption[XENON constraints with extended efficiencies]{
    XENON10 and XENON100 constraints when
    relaxing the $\Soneave \ge 1$~PE cutoff.
    Only the constraints for the fiducial (central) \Leff\ models
    are shown; from left to right, these correspond to constant,
    linearly falling, and zero \Leff\ below 3.9~keVnr.
    To illustrate the potential effect of the low recoil energy
    behavior of \Leff, these contraints are generated by arbitrarily
    assuming the nuclear recoil band cut efficiency is constant at low
    recoil energies.
    This efficiency should actually fall at sufficiently low recoil
    energies (for XENON10, at energies that yield an average \Sone\ 
    somewhere below 1~PE); these constraints should therefore not
    be taken as true constraints on the WIMP mass and cross-section.
    The actual constraints based upon a proper accounting of the
    efficiencies at low recoil energies will fall somewhere between
    the constraints shown here and in the previous figures.
    }
  \label{fig:extended}
\end{figure*}

As the potential effect of the low energy \Leff\ behavior on the
XENON100 constraints is masked by the $\Soneave \ge 1.0$~PE cutoff,
we show in \reffig{extended} the XENON100 constraints for the
three fiducial \Leff\ models when this cutoff is relaxed.
In this figure, we have arbitrarily assumed the nuclear recoil band cut
efficiency is constant at low recoil energies.  In reality, this
efficiency should fall at very low recoil energies and this
approximation becomes inappropriate at recoil energies that yield
\Soneave\ somewhere below 1~PE.  For this reason, these constraints
should not be taken to be valid constraints; we show them only to
illustrate the potential effect of low energy recoils and the low
energy \Leff\ behavior.  With all cut efficiencies properly taken
into account, the true constraints would lie somewhere between the
constraints shown in this figure and those shown in the previous
figures.

With the relaxing of the \Soneave\ cutoff, the XENON100 constraints are
still nearly identical in the DAMA and CoGeNT
regions for all three cases of low energy \Leff\ behavior.
These constraints are very similar to the ones found in the previous
figures, when a \Soneave\ 
cutoff was included, and are actually identical for the zero \Leff\ 
model as this model has no contributions from recoils with
$\Soneave < 1.0$~PE anyways.  The three
cases only begin to differ significantly in the low mass, high
cross-section parameter space located around and above the DAMA regions
in this figure.
This can be explained by the XENON100 S1 analysis threshold of
4~PE's (the full analysis range is 4-20~PE's).  In the absence of a
finite energy resolution, this corresponds to a nuclear recoil energy of
9.5~keVnr in our fiducial \Leff\ models, well into the energy range
where the \Leff\ behavior is known.  With a Poisson fluctuation in the
number of observed PE's,  recoils at lower energies have a finite
chance of producing 4 or more PE's and falling into the analysis range,
even if the average number of PE's for events at those energies is
below 4.  However, at 3.9~keVnr, the average expected number of PE's is
1.0; only 1.9\% of such events yield 4 or more PE's.  Recoils of
3~keVnr yield an average number of expected PE's of 0.79 and 0.61
for the constant and falling \Leff\ cases, respectively, with
corresponding probabilities of being observed (4+ PE's) of 0.87\% and
0.36\% (the third case, zero \Leff, produces no PE's at these
energies).
The small fraction of recoil events with energies below 3.9~keVnr
that will be observed in the analysis range means that their
contribution is only significant when there are essentially no events 
at higher energies (due to low WIMP masses and a finite escape velocity
in the halo), but to produce a sufficient number of events to fall into
the analysis range requires a very large number of WIMP scatters in
the $\sim$1-4~keVnr range, which requires a high WIMP cross-section.
Thus, even when using an overly optimistic nuclear recoil band efficiency,
the three \Leff\ cases can only result in different constraints
in the low mass, high cross-section region.
This is not necessarily the case for \Leff\ curves based on
measurements that yield values higher than Manzur \etal, as this
would push the analysis range corresponding to 4-20~PE's to lower recoil
energies; such cases, however, inevitably move the XENON100 constraints
to the left.  In any case, the most significant issue in the XENON100
analysis is the choice of \Leff\ measurements used to determine the
\Leff\ dependence, not the \Leff\ behavior at low energies.

We now turn to the XENON10 bounds.  We have reanalyzed the
XENON10 results in terms of the same \Leff\ models as used for XENON100
and discussed in the previous section; our results  differ from
those shown by the XENON collaboration due to the difference in \Leff\ 
used in their analyses and  ours.  The XENON10 results are
important because of the lower \Sone\ threshold of about 2~PE's used in
that analysis, which corresponds to 4.6~keVnr nuclear recoil energies
in our fiducial \Leff\ models (neglecting Poisson fluctuations),
much lower than the 9.5~keVnr of the XENON100 4~PE threshold.
Because of the lower threshold, the behavior of \Leff\ at low recoil
energies is relevant in producing the XENON10 constraints as Poisson
fluctuations allow for a non-trivial probability of seeing 2+~PE's for
recoil energies below 3.9~keVnr.

For the constant \Leff\ case shown in \reffig{ConstantLeff}, the lower
threshold allows for a stronger sensitivity to lower WIMP masses for
XENON10 relative to XENON100.  The fiducial case excludes at the 90\%
C.L.\ all of the CoGeNT region and DAMA to the 5$\sigma$ contour.
When the 1$\sigma$ uncertainties in the \Leff\ measurements are taken
into account, the constraints relax: DAMA is excluded to only the
3$\sigma$ contour and the some of CoGeNT region at WIMP masses below
9~GeV survive.
The XENON10 constraints mildly weaken if \Leff\ is taken to fall
linearly to zero below 3.9~keVnr, as seen in \reffig{FallingLeff}.
The fiducial case still excludes all of the CoGeNT region and DAMA to
about the 4$\sigma$ contour (not shown).  The \Leff\ 1$\sigma$ band
here allows the same DAMA and CoGeNT regions to survive as with the
constant \Leff\ case.
For the case where \Leff\ is zero below 3.9~keVnr, shown in
\reffig{ZeroLeff}, the XENON10 constraints further weaken and approach
the XENON100 constraints as the low energy events are essentially
turned off and the lower XENON10 threshold becomes less relevant.
Again, we note that this last case (zero \Leff\ at low recoil energies)
is an extremely conservative case.

As with XENON100, the potential effect of the low energy \Leff\ 
behavior on the XENON10 constraints is limited by the imposed
$\Soneave \ge 1.0$~PE cutoff.  We also show in \reffig{extended} the
XENON10 constraints for the three fiducial \Leff\ models when this
cutoff is relaxed.  The same caveats apply: the nuclear recoil band cut
efficiency that is used is not appropriate for the full recoil energy
range that it is applied over, so these do not represent valid
constraints.  Again, these constraints are only used to illustrate the
potential impact of the low energy \Leff\ behavior on XENON10
constraints.  The actual constraints when all efficiencies are
accounted for properly would fall somewhere between the constraints
shown in \reffig{extended} and those shown in
Figs.~(\ref{fig:ConstantLeff})-(\ref{fig:ZeroLeff}).

With the \Soneave\ cutoff relaxed, \reffig{extended} shows how
the low threshold allows for a strong XENON10 sensitivity to lower
WIMP masses.
This is particularly evident with the constant \Leff\ case where,
as can be seen in \reffig{S1Enr}, recoils of energy 1~keVnr yield an
average \Sone\ signal of 0.4~PE; $\sim$6\% of such recoils will
produce the necessary 2+~PE.  For the falling \Leff\ case, that
same average \Sone\ signal of 0.4~PE occurs at a higher recoil energy
of 2~keVnr, but this energy is still sufficiently low to provide
sensitivity to low mass WIMPs.
The presence of these non-trivial Poisson fluctuations at low recoil
energies leads to a very strong dependence of the XENON10 constraints
on the low energy \Leff\ behavior.
This should remain the case even when the various efficiencies are
handled properly,
though not quite to the degree shown in \reffig{extended}.
In particular, when the proper efficiencies are included, the XENON10
constraints in the constant and falling \Leff\ cases should gain
an upward curve at low WIMP masses, as seen with the other constraints,
rather than the current linear appearance.
These linear portions of the constraints at low WIMP masses
(as they appear with the logarithmic scaling of the figure)
continue to arbitrarily low WIMP masses; however, they arise from
the Poisson tails of increasingly smaller energy events that would be
suppressed when using the proper efficiencies.

Though we have not included it in this work, CDMS Silicon data may
provide further constraints on the DAMA and CoGeNT regions and should
be considered in a full discussion of compatibility between the
various experimental results.  The CDMS Silicon results will be
considered in future work.

In summary, we have examined a number of subtleties relevant to direct
detection studies of low mass WIMPs.
In the interest of examining the most conservative XENON constraints,
we have used the Manzur \etal\ \cite{Manzur:2009hp} data alone in our
analyses. Of the existing data sets, the Manzur \etal\ data yield the
lowest values for \Leff, implying higher recoil energy thresholds for
the XENON experiments, and thereby reducing the sensitivity of XENON
to low mass WIMPs.
We find that, when basing the \Leff\ curves on these Manzur \etal\ 
measurements, the behavior of \Leff\ at low energies (less than
3.9~keVnr) has negligible effect on the XENON100 constraints in the
regions of interest for DAMA and/or CoGeNT.
For XENON100, the choice of data sets upon which the \Leff\ dependence
is based is more important than the extrapolated behavior of \Leff\ at
low recoil energies.
The strongest bounds are from XENON10, rather than XENON100, due to the
lower energy threshold.
For reasonable choices of \Leff\ and for the case of spin independent
elastic scattering, we find that
XENON10 is incompatible with the DAMA/LIBRA 3$\sigma$ region
and severely constrains the CoGeNT 7-12~GeV WIMP mass region.


\begin{acknowledgments}
  C.S.\ is grateful for financial support from the Swedish Research
  Council (VR) through the Oskar Klein Centre.
  G.G.\ was supported in part by the US Department of Energy Grant
  DE-FG03-91ER40662, Task C.
  P.G.\ was  supported  in part by  the NFS grant PHY-0456825 at the
  University of Utah.
  K.F.\ acknowledges the support of the DOE and the Michigan Center for
  Theoretical Physics via the University of Michgian. 
  G.G., P.G., and C.S.\ thank the Galileo Galilei Institute for
  Theoretical Physics for the hospitality and the INFN for partial
  support during the completion of this work.
  We thank also E.\ Aprile, K.\ Arisaka, D. Hooper, and K. Zurek for
  helpful conversations;
  P.\ Sorensen for bringing to our attention the XENON10
  \Sone\ peak finding efficiency factor that was missing in the
  first version of this paper;
  L.\ Baudis, A.\ Manalaysay, G.\ Plante, and P.\ Sorensen for
  discussions regarding the XENON detectors and analysis;
  and B.\ Edwards and T.\ Sumner for discussions regarding the
  ZEPLIN-III \Leff\ estimates.
\end{acknowledgments}


\appendix*

\section{\label{sec:Response} Comment about our choice of relevant
         parameters and response to critique}

Shortly after the first version of our paper was released,
Ref.~\cite{Collar:2010nx} appeared commenting on it.
We include here a detailed explanation of the XENON10 efficiencies and
cuts we are using, as well of our use of the Manzur \etal\ measurements
as a conservative choice for \Leff\ (instead of the ZEPLIN-III
measurements) which we believe are relevant in view of the comments
expressed in Ref.~\cite{Collar:2010nx}.

\subsection{\label{sec:etaS1} XENON10 efficiencies and cuts}

The \Sone\ peak finding efficiency factor \etaSone\ had not been
included in the XENON10 analysis in the first version of our paper
(as correctly pointed out in Ref.~\cite{Collar:2010nx})
and has been accounted for in this revised version.
However, this leads to only a moderate weakening of the XENON
constraints by shifting the bound on the cross section upward by less
than a factor of two.  It is important to point out that including this
effect does not weaken the constraints upwards in cross-section by
2-3 orders of magnitude, contrary to the claim in the critique given in
Ref.~\cite{Collar:2010nx}.  One can understand the small magnitude of
the effect with the following reasoning.

A valid signal event in the XENON detectors is required to produce
coincident scintillation in at least two PMTs.
The \etaSone\ factor accounts for experimental limitations in
identifying and reconstructing at least two PMT contributions to
the overall \Sone\ signal of a recoil event.
The \Sone\ signal is determined from the area under the peaks produced
in the electronic readout of the PMTs.
Due to digitization of the signal and intrinsic PMT performance, the
size and shape of the peaks will vary\footnote{
    The measured \Sone\ signal from a single PE as determined from the
    area of these peaks is $1.0 \pm 0.6$~PE.  Thus, \Sone\ can take
    on non-integer values even though it is given in terms of number
    of PE.
    };
see Fig.~14 of Ref.~\cite{Aprile:2010bt} for an example.
Small or poorly shaped peaks may fail to be properly tagged as a PE
peak in a PMT.  Using the more conservative estimate found in
Ref.~\cite{Sorensen:2010HEFTI}, only $\etaSone \approx 60\%$ of 2~PE
events will have both PE peaks properly tagged and identified as
coincident.  As the number of PE's increases, the probability of
passing the two-fold PMT requirement rapidly rises to 100\%.

The significance of the \etaSone\ factor on XENON10 constraints can
be easily estimated.  If this factor is conservatively assumed to be
60\% over the entire \Sone\ analysis range (2+~PE) instead of just at
low \Sone, the expected number of events passing the XENON10 cuts at
all energies should fall to 60\% of the number of events expected
without \etaSone\ applied.  As the recoil rate at a particular WIMP
mass is proportional to the scattering cross-section, the original
recoil spectrum can be exactly reproduced by shifting the cross-section
upwards by a factor of $1/0.6 \approx 1.7$.  Thus, if a WIMP mass and
cross-section is excluded by XENON10 at some given CL without the
\etaSone\ factor applied, a WIMP with the same mass and $\times$1.7
higher cross-section would yield the same excluded spectrum with
the \etaSone\ factor included.  In this case, the XENON10 exclusion
curves shift upwards in cross-section by $\times$1.7.  As \etaSone\ 
is higher than 60\% for \Sone\ larger than 2~PE, the shift in
cross-section will be even smaller at heavier WIMPs where a significant
number of high \Sone\ events are expected.

\begin{figure*}
  \insertwidefig{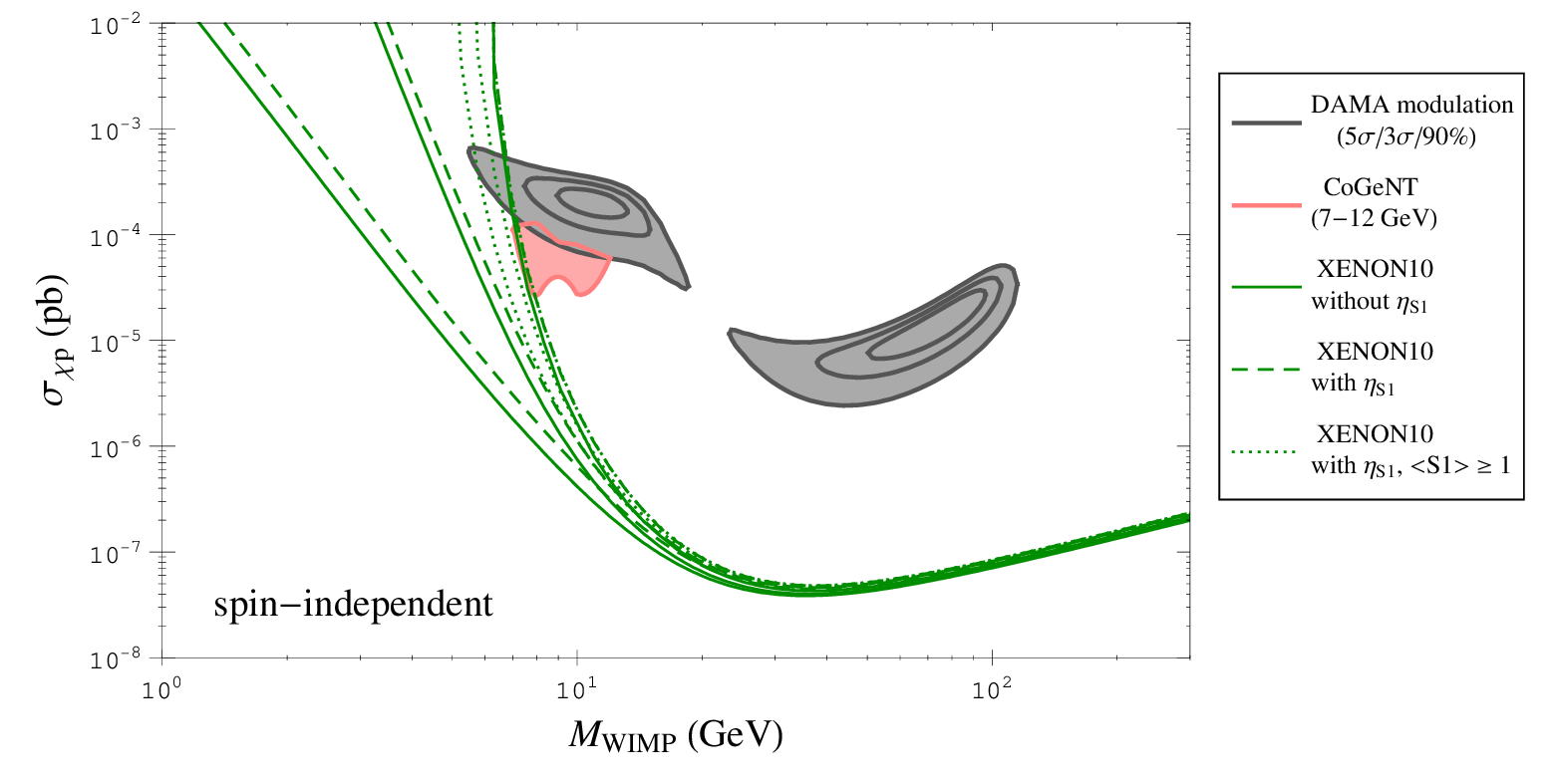}
  \caption[\etaSone\ and \Sone\ cutoff]{
    Impact of
    (i) the \Sone\ peak finding efficiency \etaSone\ (dashed lines) and
    (ii) the $\Soneave \ge 1$ cutoff (dotted lines)
    on the XENON10 constraints.
    Only the constraints for the fiducial (central) \Leff\ models
    are shown; from left to right, these correspond to constant,
    linearly falling, and zero \Leff\ below 3.9~keVnr.
    One can see that the second effect is by far the stronger one.
    }
  \label{fig:etaS1}
\end{figure*}

We demonstrate in \reffig{etaS1} the weakening of the XENON10
constraints when \etaSone\ is included.  We show only the three
fiducial (central) \Leff\ models described in our paper; constraints
without and with \etaSone\ included are given by solid and dashed
curves, respectively.  For a given WIMP mass, the constraint
increases by $\sim$ $\times$1.7 in the scattering cross-section at
low WIMP masses and by less at higher WIMP masses.  This is a
relatively modest weakening of the constraint given the logarithmic
scaling of the exclusion curve figures. 
Please recall that we have used the more conservative estimate
of \etaSone\ found in Ref.~\cite{Sorensen:2010HEFTI}.  Had we used the
\etaSone\ factor given in Ref.~\cite{Sorensen:2008ec}, the change would
have been even milder.

We will clarify an issue that is perhaps the source of what appears to
be an erroneous application of the \etaSone\ factor in
Ref.~\cite{Collar:2010nx} that yielded a greater weakening in the
XENON10 results than expected.
Given a recoil at some energy \Enr, \refeqn{S1} gives the
\textit{average} expected \Sone\ scintillation signal \Soneave.
Given the measured \Sone\ in an event, \refeqn{S1} can be inverted to
obtain a reconstructed recoil energy \Eest.
Due to the discreteness of the produced PE's and the variation in the
\Sone\ peaks,the observed prompt scintillation signal \Sone\ is
a random value whose expectation value is \Soneave.  Thus, the
reconstructed recoil energy \Eest\ is also a random value and provides
only an estimate of the (unknown) true recoil energy, \Enr.
The distinction between \Eest\ 
and \Enr\ is important.  With the PE fluctuations and \Sone\ peak
variations, the true recoil energy cannot be precisely determined for
any given recoil event.  \Eest\ is simply an estimate of the likely
recoil energy that produced an event and is understood to have some
intrinsic uncertainty attached to it.  At high recoil energies,
$\Eest \approx \Enr$ and \Eest\ can be taken as a good approximation of
the actual recoil energy.  At low recoil energies, \Eest\ may differ
significantly from \Enr\ on an event by event basis.

The \etaSone\ factor is sometimes given as a function of recoil energy.
However, as \etaSone\ is really a function of the \Sone\ signal,
this recoil energy refers to the reconstructed energy \Eest\ of an
event (recall \Eest\ and \Sone\ have a 1:1 mapping through
\refeqn{S1}) and does \textit{not} refer to the actual recoil energy
\Enr.
Neglecting the \Sone\ peak variations and efficiencies other than
\etaSone, the fraction of events $f$ that exceed the 2~PE threshold
from scatters at a recoil energy that yields an average \Sone\ signal
\Soneave\ is given by:
\begin{equation} \label{eqn:f}
  f(\Soneave) =
    \etaSone(2) P(2|\Soneave) +
    \etaSone(3) P(3|\Soneave) +
    \etaSone(4) P(4|\Soneave) \ldots \ ,
\end{equation}
where \etaSone\ is taken as a function of \Sone\ and $P(k|\Soneave)$ is
the Poisson probability of seeing $k$~PE with an average of $\Soneave$.
If the \etaSone\ factor were assumed to be a function of the actual
recoil energy \Enr, then
\begin{equation} \label{eqn:fbad}
  f(\Soneave) =
    \etaSone(\Soneave)
    \left[
      P(2|\Soneave) + P(3|\Soneave) + P(4|\Soneave) \ldots
    \right]
\end{equation}
would erroneously be taken as the fraction of events exceeding the
\Sone\ threshold.  Use of \refeqn{fbad} in place of \refeqn{f}
will improperly yield greatly weakened XENON10 constraints at low
WIMP masses.

A second change has been made in the present version of this paper.
In the first version, we assumed the same nuclear recoil
band efficiency given for $2 \le \Sone \le 5$ in
Ref.~\cite{Angle:2009xb} applied to all Poisson fluctuated events that
appeared in that range.  This was an optimistic assumption that may
overestimate the XENON10 sensitivity at low WIMP masses.  There are two
data cuts relevant for low energy recoils that yield
$2 \le \Sone \le 5$: a recoil event must have $\Stwo \gae 300$~PE
(\Stwo\ threshold) and $1.88 \le \log_{10}(\Stwo/\Sone) \le 2.40$
(nuclear recoil band) to be accepted as a valid event.
The latter requirement is a cut designed to exclude electron recoil
background events which tend to produce higher values of $\Stwo/\Sone$
than nuclear recoils; the range of values here accept the lower
$\sim$45-50\% of the distribution of $\Stwo/\Sone$ expected for
nuclear recoils, as determined from calibration data.
At very low recoil energies, an event that produces an upward
fluctuated $\Sone \ge 2$~PE and would otherwise be considered a valid
event might fail to produce enough \Stwo\ signal to pass the \Stwo\ 
threshold.  The efficiency can thus be expected to fall at low
recoil energies.
To avoid problems with these two data cuts, we have added a cutoff to
our analysis and ignore recoil energies for which $\Soneave < 1.0$~PE.
We also take the same 47\% nuclear recoil band efficiency in the
2-5~PE \Sone\ bin for all events in this \Sone\ range stemming from
low energy recoils.
The addition of this cutoff significantly weakens the XENON10
constraints for WIMP masses below $\sim$6~GeV for the constant and
falling \Leff\ models discussed in the paper, as seen by the dotted
curves in \reffig{etaS1}.  This cutoff has a far greater impact on the
constraints than the inclusion of the \etaSone\ factor and is the
dominant source of the change in XENON10 constraints from the first
version of this paper.  The third \Leff\ model, zero below recoil
energies of 3.9~keVnr, is unaffected by this cutoff as it does not yield
any observable events at low energy anyways.

The $\Soneave \ge 1.0$~PE cutoff and 47\% assumed efficiency we have
adopted here are conservative for two reasons:
(1) The \Sone\ and \Stwo\ fluctuations are independent.  This means
that the Poisson fluctuated events that yield higher \Sone\ than the
average will have lower $\Stwo/\Sone$ ratios on average and are more
likely to pass the nuclear recoil band cut.  The upward Poisson
fluctuated events for $1.0 < \Soneave < 2.0$ can survive this cut
as much as 70-80\% of the time, higher than the assumed 47\%.
(2) Our choice of a cutoff at $\Soneave = 1$~PE is due to our limited
ability to examine the efficiencies at lower recoil energies, not
due to an expected lack of events at these low energies.  We note
that there will also be fluctuations in the number of ionization
electrons that lead to the \Stwo\ signal\footnote{
    The \Stwo\ threshold of $\sim$300~PE corresponds to about
    12 ionization electrons, each of which produce $\sim$25~PE.%
    };
even at very low recoil energies, there may be a non-zero probability
of both \Sone\ and \Stwo\ exceeding their respective thresholds and
producing events in the analysis region that pass all cuts.
For the constant \Leff\ case at a WIMP mass of 5~GeV and scattering
cross-section of 10$^{-2}$~pb, a point not excluded by the
conservative bounds in \reffig{etaS1}, about 20,000 recoils would
be expected for recoil energies corresponding to
$0.5 \le \Soneave \le 1.0$.  Even with a very small efficiency for
such recoils, some number of these events would be expected to have
fluctuations that put them in the XENON10 analysis region.
If the nuclear recoil band efficiency were only $\sim$1\% over this
range, 10+ events would be expected to pass the various cuts;
the lack of such events in the data would rule out these WIMP
parameters.

\subsection{\label{sec:ZEPLINLeff} ZEPLIN-III \Leff\ models}

Refs.~\cite{Collar:2010gg} and \cite{Collar:2010nx} suggest using the
the ZEPLIN-III \Leff\ measurement \cite{Lebedenko:2008gb}, represented
as a band in Fig.~1 of Ref.~\cite{Collar:2010gg}, as a conservative
choice of \Leff. Thus, we discuss here this possibility.

ZEPLIN fits a nonlinear \Leff\ model
to their broad spectrum nuclear recoil calibration data to obtain
\Leff\ curves that were used in their analysis.  These fits suggest a
constant \Leff\ at recoil energies above $\sim$30~keVnr, with \Leff\ 
sharply falling at energies below $\sim$20~keVnr and approaching zero
at $\sim$7-8~keVnr; see Figure~15 of Ref.~\cite{Lebedenko:2008gb} and
the accompanying text.  
The suggestion has been made \cite{Collar:2010nx} that \Leff\ should be
taken to be zero below $\sim$8~keVnr as a conservative model of \Leff.
This \Leff\ model would yield significantly weaker XENON constraints
relative to what we have referred to as conservative models based on
the Manzur \etal\ measurements of \Leff\ \cite{Manzur:2009hp}.

We have several reservations about using this ZEPLIN inspired
\Leff\ model.  First, ZEPLIN does not provide estimates of \Leff\ below
$\sim$7~keVnr and, in fact, states they are limited in constraining
\Leff\ at these low energies
(in the caption of their Fig~15, it is said that
``The constraints become very weak outside the energy ranges shown.'').
Second, neither the technical details of ZEPLIN's curve-fitting
nor estimates of the statistical and systematic uncertainties are
provided in their paper.
Without these, it is unclear with which degree of certainty the low
energy ($\sim$7-10~keVnr) end of the ZEPLIN-III curves
should be treated.
Several sources of error may contribute to the level of uncertainty
in these \Leff\ estimates.
One is the errors associated with the Monte Carlo used to compare with
data, which are difficult to quantify since they can arise from
inaccuracies in the inelastic neutron scattering data used in the
G{\'e}ant4 code and obtained from an international library of such data.
Another issue are systematic errors associated with uncertainties in
the position of the neutron source near the detector during calibration
runs.
If the uncertainties are large, then ZEPLIN does not have the
statistical power to determine the \Leff\ behavior at low recoil
energies and one should base the \Leff\ curves on measurements that
have a better statistical power to analyze this \Leff\ behavior, such
as the Manzur \etal\ data.
If the uncertainties are small so that the ZEPLIN curve is expected to
be an accurate representation of the \Leff\ behavior at low energies,
then there is a very serious discrepancy between the ZEPLIN and
Manzur \etal\ results (see below).
The ZEPLIN collaboration claims no leverage on \Leff\ below about
8~keVnr and do not suggest that it goes to zero in that energy
range\footnote{
    ZEPLIN used a 9-point spline function to model the curves and a
    classical maxmium likelihood method using a grid scan across a
    9~point parameter space to make sure they found a global minimum
    \cite{Sumner:2010pc}.  ZEPLIN only fit their data down to 2~keVee,
    thus could infer nothing below about 6~keVnr. The middle curve in
    their Fig.~15 \cite{Lebedenko:2008gb} gives the ``best fit'' and
    the outer two curves indicate regions where ``similar
    goodness-of-fit values'' were obtained as stated in the caption;
    the band does not represent a region with a particular statistical
    significance (\eg\ 1-$\sigma$ or 90\% C.L.) in the uncertainties.
    }~\cite{Sumner:2010pc}.
The recoil energy interval in the ZEPLIN analysis
of 10.7--30.2~keVnr does not include the low energy,
low \Leff\ end of their \Leff\ curves.  In this analysis region, the
ZEPLIN \Leff\ curves are compatible with the Manzur \etal\ 
measurements within the 1-2$\sigma$ level.

Furthermore, the ZEPLIN inspired model proposed by
Ref.~\cite{Collar:2010nx},
taking $\Leff = 0$ below recoil energies of $\sim$8~keVnr,
is strongly incompatible with the Manzur \etal\ measurements
as well as other fixed energy neutron scattering measurements
such as those of Aprile \etal\ \cite{Aprile:2008rc}.
We reiterate that our choice to use the Manzur \etal\ data in our
analysis was due to the lower \Leff\ values that give more conservative
XENON constraints and should not be construed as an indication that
we regard this data as the most accurate available.  We make no
contribution to the discussion regarding the accuracy of the various
low energy recoil measurements from neutron beam scattering.  See
Ref.~\cite{Manalaysay:2010mb} for a discussion of the various potential
issues which may affect the \Leff\ determinations in these experiments.
In any case, much of the discussion below regarding the incompatibility
of the Manzur \etal\ data with $\Leff = 0$ applies also to the
Aprile \etal\ measurements.

Given the Manzur \etal\ measurement of \eg\ 
$\Leff = 0.073_{-0.025-0.026}^{+0.034+0.018}$ (statistical and
systematic errors, respectively) at a recoil energy of
$3.9 \pm 0.9$~keVnr, one might naively conclude this measurement is
consistent with $\Leff = 0$ at the $\sim$2-3$\sigma$ level.
However, the \Leff\ value and errors are determined from a
$\chi^2$ fit to the data as a function of \Leff.  The $\chi^2$ versus
\Leff\ curves, such as shown in Fig.~11(c) of
Ref.~\cite{Manzur:2009hp} for a recoil energy of 6~keVnr, are not
symmetric about the minima (which provide the central values), but rise
very rapidly as \Leff\ becomes small, effectively diverging at
$\Leff = 0$.  The result is that the Manzur \etal\ measurements
(four of which are below recoil energies of 7~keVnr) are incompatible
with \Leff\ $=0$ recoil energies of $\sim$8~keVnr, as this would yield
$\chi^2$ values that are imcompatible at far higher than the 3$\sigma$
level.  We consider the Manzur \etal\ measurements more reliable than
the ZEPLIN-III estimate of \Leff\ at low energies.

To illustrate the incompatibility of the Manzur \etal\ data with
$\Leff = 0$, one can understand the $\chi^2$ behavior  as follows.
Manzur \etal\ measured \Leff\ at several
recoil energies by observing the scintillation response in a Xenon
detector for neutrons from a beam of fixed energy (2.8~MeV) that
scatter in a fixed direction; the recoil energy is fixed by the angle
of scatter\footnote{
    Due to the geometry of the Xenon detector and width of the
    scintillator used to detect the scattered neutrons, each
    measurement is sensitive to a finite range of recoil angles
    and, thus, recoil energies.  The uncertainties in the recoil
    energies given for each \Leff\ measurement arise mainly from this
    range.%
    }.
If \Leff\ is non-zero at the recoil energy corresponding to a
particular scattering angle, a histogram of the \Sone\ from observed
events will generate a peak due to scintillation from the single
scatter events in the detector.
The primary background is double scattering neutrons which can reach
the neutron scintillator located at the fixed angle from the neutron
beam, but scatter with a different energy than expected for single
scatter neutrons that reach the neutron scintillator; one of the
two scatters must occur outside of the active volume for the event
to be mistaken as a single scatter.  These double
scattering neutron events are expected to occur at an almost negligible
rate compared to the single scatter signal events.  In addition, these
double scatter events produce a fairly flat \Sone\ distribution, not
a peak (see \eg\ Figure~9 of Ref.~\cite{Manzur:2009hp}).
Figures~11(a) and 11(c) of Ref.~\cite{Manzur:2009hp} show an \Sone\ 
histogram and corresponding $\chi^2$ versus \Leff\ curve, respectively,
for recoils centered at 6~keVnr.  The histogram shows a peak of
$\sim$700 events (consistent with scintillation from single scatter
signal events) with a negligible flat contribution (expected from any
significant double scatter background).  The $\chi^2$ in this case
is minimized at $\Leff \approx 0.06$.  At lower \Leff, the $\chi^2$
grows rapidly for this reason: it is difficult to account for this
$\sim$700 scintillation event peak if the nuclear recoils produce
little to no scintillation.  Note the systematic issues discussed in
Manzur \etal\ generally affect the interpretation of which \Leff\ 
would produce such a peak, but do not provide significant alternate
mechanisms for producing such a peak aside from the scintillation from
low energy recoils (scintillation that can only be produced if
$\Leff \not\approx 0$).
Thus, while determining the value of \Leff\ based on the size and
shape of such \Sone\ peaks may be limited by statistics or
biased by systematic effects, the simple presence of such a
peak is strong evidence that \Leff\ is non-zero.

Given that Manzur \etal\ have observed scintillation peaks at multiple
recoil energies over $\sim$ 4-8 keVnr, we consider the ``conservative''
\Leff\ model suggested in Refs.~\cite{Collar:2010gg} and
\cite{Collar:2010nx}, with $\Leff = 0$ below $\sim$8~keVnr,
to be grossly incompatible with existing data and therefore
unrealistically conservative.


\end{document}